\documentclass[aps,prd,preprint,groupedaddress,showpacs,showkeys,twocolumn,10pt]{revtex4-2}
\usepackage{slashed} 
\usepackage{amsmath} 
\usepackage{longtable}
\usepackage{tabularx}
\usepackage{color}
\usepackage{graphicx}
\usepackage[caption=false]{subfig}
\usepackage{lineno}
\usepackage{epsfig}
\usepackage{hyperref}   

\begin{document}
	\title{Calibration of LHAASO-WCDA Energy Scale using Cosmic Ray Moon shadow}
	\date{\today}

\author{    
Zhen Cao$^{1,2,3}$,
F. Aharonian$^{3,4,5,6}$,
Y.X. Bai$^{1,3}$,
Y.W. Bao$^{7}$,
D. Bastieri$^{8}$,
X.J. Bi$^{1,2,3}$,
Y.J. Bi$^{1,3}$,
W. Bian$^{7}$,
J. Blunier$^{9}$,
A.V. Butkevich$^{10}$
% ...
} 

\affiliation{
$^{1}$ State Key Laboratory of Particle Astrophysics \& Experimental Physics Division \& Computing Center, Institute of High Energy Physics, Chinese Academy of Sciences, 100049 Beijing, China\\
$^{2}$ University of Chinese Academy of Sciences, 100049 Beijing, China\\
$^{3}$ Tianfu Cosmic Ray Research Center, 610000 Chengdu, Sichuan,  China\\
$^{4}$ University of Science and Technology of China, 230026 Hefei, Anhui, China\\
$^{5}$ Yerevan State University, 1 Alek Manukyan Street, Yerevan 0025, Armenia \\
$^{6}$ Max-Planck-Institut for Nuclear Physics, P.O. Box 103980, 69029  Heidelberg, Germany\\
$^{7}$ Tsung-Dao Lee Institute \& School of Physics and Astronomy, Shanghai Jiao Tong University, 200240 Shanghai, China\\
$^{8}$ Center for Astrophysics, Guangzhou University, 510006 Guangzhou, Guangdong, China\\
$^{9}$ APC, Universit'e Paris Cit'e, CNRS/IN2P3, CEA/IRFU, Observatoire de Paris, 119 75205 Paris, France\\
$^{10}$ Institute for Nuclear Research of Russian Academy of Sciences, 117312 Moscow, Russia\\
% ...
}


\date{\today}
\email[]{saeed.physics@pu.edu.pk}
\email[]{faisal.chep@pu.edu.pk} 
\email[]{caozh@ihep.ac.cn}
\thanks{}	
    
	\begin{abstract}
	Cosmic-ray particles in the TeV range, originating outside the solar system, are partially blocked on their way to Earth by the Moon, producing a deficit in their flux known as the Moon shadow. This work presents a precision measurement of the westward shift of the Moon's shadow using 3.4 years of data from LHAASO's Water Cherenkov Detector Array (WCDA). We use this direct physical measurement, combined with four cosmic-ray composition models, each divided into five mass groups, to calibrate the detector’s absolute energy response in the range of 4 - 37 TeV. We quantify the fractional deviation between the calibrated and simulated energy scales by determining a best-fit scale shift parameter of $\epsilon = -0.04 \pm  0.03$. This yields a 95\% confidence interval of -9\% to +0.2\% for the systematic energy scale deviation. Complementing the simulation based on hadronic interaction model, this geomagnetic calibration provides a data-driven physical validation of the simulated WCDA energy response, effectively bounding systematic uncertainties and refining the overall energy scale.
			
	\end{abstract}
	
	\maketitle
\section{Introduction}	
	The cosmic ray Moon shadow refers to a reduction in the observed cosmic ray flux in the direction of the Moon. As the Moon obstructs incoming cosmic ray particles, detectors located on Earth or in space measure a diminished cosmic ray flux in its vicinity, resulting in a “shadow” effect. The shadow is, however, slightly displaced westward due to the deflection of primary particles by the Earth’s magnetic field. Since the deflection is energy dependent, the phenomenon can be used to calibrate the energy scale of cosmic ray detectors along with estimating their angular resolution. It is particularly important for ground-based detectors, in which the atmosphere acts as a calorimeter, and only a fraction of secondary cosmic ray particles produced by a primary through an extensive air shower (EAS) are detected. Energy calibration is therefore an essential aspect of cosmic ray physics with ground-based detectors, playing a crucial role in precise measurements of energies of primary cosmic ray particles.   
    
	\par 
    The first experimental observation of the Moon and Sun shadow of cosmic-rays, which achieved the significance of 4.9$\sigma$ and an angular resolution of $0.75_{-0.09}^{+0.13}$, was reported in 1991 ~\cite{AA91}. The ARTEMIS (Antimatter Research Through The Earth Moon Ion Spectrometer) experiment used its Imaging Atmospheric Cerenkov telescope to search for the Moon's shadow in TeV cosmic rays; however, atmospheric fluctuations constrained its sensitivity, preventing detection of a percent-level antimatter component ~\cite{PBU01}. Using the Soudan 2 experiment with a depth of 2090 mwe (meter water equivalent), the Moon’s shadow was observed in the underground muon flux at a significance of 5$\sigma$, revealing an angular resolution better than 0.3 degrees and confirming the detector alignment stability within 0.15 degrees over ten years ~\cite{COBB02}. The MACRO experiment observed the Moon shadow with 6.5 $\sigma$ significance and the Sun shadow with 4.6$\sigma$ significance, and established an upper limit of 48\% on the antimatter flux in cosmic rays around 10 TeV, using muon flux data collected from 1989 to 2000 ~\cite{AMB03}. Using the L3+C muon spectrometer at CERN, the shadowing effect of the Moon of cosmic rays was observed with 9.4$\sigma$ significance, revealing a strong geomagnetic field (GMF) effect and setting an upper limit of 0.11 on the antiproton-to-proton ratio with 90\%C.L. at 1 TeV  ~\cite{L307}. With the Tibet-III air shower array, the Moon’s shadow was observed with 40$\sigma$ significance, exhibiting a 0.23 degrees westward shift caused by GMF, and providing a 90\%C.L. upper limit of 7\% on the multi-TeV anti-proton-to-proton flux ratio ~\cite{AM07}. The Main Injector Neutrino Oscillation Search (MINOS) detector observed the Moon's shadow at 5.6$\sigma$ and the Sun's shadow at 3.8$\sigma$, establishing an absolute pointing of $0.17\pm0.12^\circ$ and revealing hints of interplanetary magnetic field effects in both shadows ~\cite{MINOS11}. 
    
    \par The ARGO-YBJ experiment observed the cosmic-ray Moon shadow effect in the multi-TeV energy region with 55$\sigma$ significance, using its westward shift from the GMF to calibrate the primary energy scale and its shape to determine the angular resolution and pointing accuracy. Furthermore, the shadow was studied to investigate the effects of the solar wind through day/night comparisons and demonstrated as a comprehensive tool for characterizing the performance of an air-shower array ~\cite{BBB11}. In addition, three years of IceCube observations further establish the cosmic-ray Moon and Sun shadows as high-significance calibration tools, with both shadows detected at greater than 12$\sigma$ across different detector configurations ~\cite{IC15}. Extending this Moon-shadow study, ARGO-YBJ disentangle composition dependence in absolute-energy scale calibration using Moon-shadow data dominated by the light component, yielding a 17\% energy scale error from 3 TeV to 50 TeV and validating the technique against the light component cosmic-ray spectrum over the same energy range \cite{BBB17}. LHAASO-KM2A has recently published cosmic-ray Moon shadow measurements that precisely confirm its Monte Carlo detector energy response over the 20-260 TeV range, yielding an energy scale uncertainty of 14\% to 17\% at a confidence level of 95\% ~\cite{ali24}.
    \par 
	Cosmic ray air shower experiments like the Tibet AS$\gamma$, HAWC (High Altitude Water Cherenkov Observatory) and LHAASO (Large High Altitude Air Shower Observatory) are designed to observe a cascade of secondary cosmic ray particles using an array of ground-based detectors. These experiments use different detection techniques, including water Cherenkov detectors, scintillation counter, and air Cherenkov telescopes, to measure the arrival times, directions, and energies of cosmic ray showers. A Moon shadow study was carried out using LHAASO WCDA data from March 2021 to December 2022, in which cosmic-ray Moon shadows were analyzed for events with compactness C $<$ 10 and C $>$ 10. The results reveal that shadow shifts decrease and shadows become more compact with increasing energy, providing support for detector calibration, event selection, and understanding of cosmic-ray trajectories in the GMF ~\cite{wcda23}. The absolute energy scale of primary cosmic rays observed by the Wide Field-of-View Cherenkov Telescope Array (WFCTA) is calibrated using the Moon’s westward shift measured for nine months of data by WCDA-I of LHAASO and propagated to WFCTA through commonly triggered events, producing consistent median energies around 21-23 TeV ~\cite{wyj21}. The present work builds directly on the previous single pond calibration (WCDA-I) 
    Although that work was limited to eight months of data, this analysis extends the data set to 3.4 years and incorporates all three WCDA ponds (WCDA-I,  WCDA-II and WCDA-III). The primary focus remains the observation of the Moon shadow using LHAASO-WCDA and its application to calibrate the detector’s energy scale, but now with significantly increased statistics and temporal coverage.
 
    \par The paper is organized as follows. The LHAASO experiment is briefly described in Section II. Section III describes the Monte Carlo (MC) simulation of air showers and detector response, the ray-tracing calculation, and the Moon shadow measurement. The determination of the energy scale is presented in Section IV. In Section V, a short conclusion is drawn.	

    
	\section{LHAASO}
\begin{figure*}[htb]
		\begin{center}
			\includegraphics[width=0.6\linewidth]{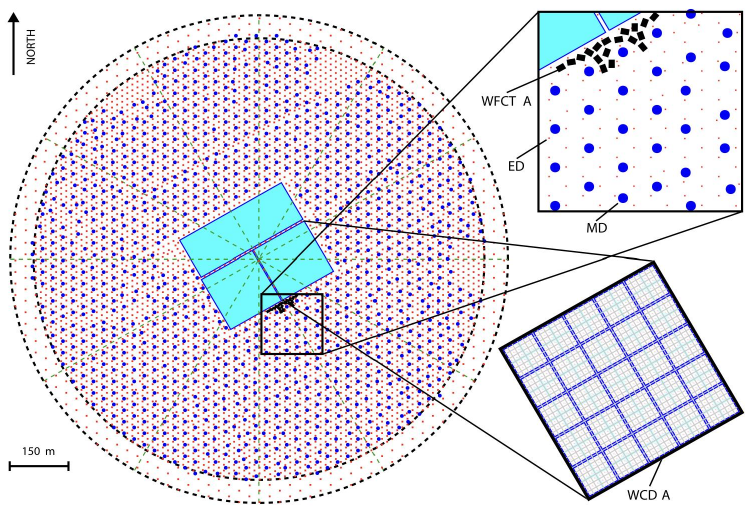}
		\end{center}
		\caption[LHAASO]{Schematic layout of LHAASO experiment, representing the arrangment of its main detector components}
		\label{1-Lhaaso-view}
	\end{figure*}

	The Large High Altitude Air Shower Observatory is a multi-array facility designed primarily for research in cosmic ray physics and gamma-ray astronomy. It is located 4410 m above sea level near the Daocheng village, in Sichuan province, P. R. China, with coordinates $100.01^\circ$E and $29. 35^\circ$N (FIG. ~\Ref{1-Lhaaso-view}). It consists of three interconnected detector arrays, including a Kilometer Square Area (KM2A) covering an area of 1.3 $\text{km}^2$, a 78,000 $\text{m}^2$ Water Cherenkov Detector Array (WCDA), and a Wide Field-of-view Cherenkov Telescope Array (WFCTA) with 18 telescopes ~\cite{CAO23}. The KM2A, the largest sub-array of LHAASO, uses 5216 electromagnetic detectors (EDs) and 1188 muon detectors (MDs) to reconstruct cosmic-ray and gamma-ray showers, with a three-level quality control system ensuring data reliability ~\cite{ZCB25}. Each ED consists of 4$\times$4 plastic scintillation tiles, each 250 mm $\times$ 250 mm $\times$ 20 mm (called 'ED unit'), packed in a cover frame with an area of 1.2 m $\times$ 1 m. In contrast, each MD is a 6.8 m diameter and 1.2 m height concrete tank, filled with pure water contained in a sealed cover with a reflective inner surface, with a PMT residing at the top of the unit ~\cite{hhh18}. The LHAASO-WFCTA, is specially designed to study the interactions of primary cosmic rays in the atmosphere along with the Cherenkov and fluorescence light. Each telescope consists of a spherical light collector of about 4.7 m$^2$ and a focal plane camera of $32 \times 32$ pixels with a pixel size of 0.5 degrees ~\cite{MA22}. 

	LHAASO's primary objective is to conduct an in-depth analysis of cosmic rays and high-energy gamma ray sources, including both galactic and extragalactic. By leveraging the unique advantages of high-altitude operations, LHAASO is committed to addressing a wide spectrum of challenges within the realm of astrophysical research.

	\subsection{WCDA}
	The LHAASO-WCDA consists of three water ponds: WCDA-I and WCDA-II (each 150 m $\times$ 150 m), and WCDA-III (300 m $\times$ 110 m). The ponds are divided into 3120 detector cells (900, 900, and 1320 units, respectively). They contain 350 ktons of purified water and cover a total area of 78,000 m$^2$.  A closed recycling system is implemented to maintain water purity and achieve an attenuation length for near-ultraviolet light longer than 15 m. Each detector cell is 5 m $\times$ 5 m, separated by non-reflecting black plastic curtains. In WCDA-I, each cell is equiped with two upward-facing PMTs (8-inch and 1.5-inch PMT combination) residing on the bottom at the center of each unit, whereas WCDA-II and WCDA-III employ a 20-inch and 3-inch PMT combination. For each PMT, a hit is formed with the threshold of 1/3 photoelectron (PE) for an 8-inch PMT, and 1 PE for a 20-inch MCP-PMT. A trigger algorithm was implemented to record cosmic ray showers by requiring at least 30 PMTs fired simultaneously in a $12 \times 12$ PMT array within a window of 250 ns. To ensure a reliable data sample, data quality checks are performed regularly based on the status of DAQ data and the quality of reconstruction ~\cite{wcda21}.

	\section{Moon shadow Analysis}\label{sec:selection}
		The observation of the Moon shadow effect provides a reliable and effective approach for calibrating the energy scale of ground-based air shower detectors like LHAASO-WCDA. The Moon blocks cosmic rays from its direction, resulting in a detectable deficit of particles in cosmic ray flux (called the Moon shadow). One of the key features used in this context is the westward displacement of the Moon's shadow center caused by the GMF. The idea of observing the Moon's shadow in cosmic rays was first proposed by G. W. Clark in 1957~\cite{GW57}.
       
		\par The Moon shadow shift for each energy bin reflects the average deflection experienced by all cosmic ray particles coming from the Moon's direction within that bin. Since the deflection depends on both the particle's energy and charge, the measured shift in each range is not a direct measure of energy; it incorporates a composition-dependent factor.

	\subsection{Monte Carlo Simulation} 
        To model the propagation of cosmic rays through the Earth–Moon system, we carried out a comprehensive Monte Carlo simulation. Air-shower development was simulated using the CORSIKA v76400 package ~\cite{HKC98}. We employ GHEISHA ~\cite{JB03} and the quark-gluon-string model QGSJET-II-04 ~\cite{KK17} to describe hadronic interactions at low and high energies, respectively. We simulated 26 primary nuclei (proton to iron) following an E$^{-2}$ differential spectrum across an energy range of 1.1 GeV to 3 PeV, assuming an isotropic arrival direction distribution. Using the knee-free Hörandel spectrum in each segment, we re-weighted the events to recover the spectrum for arbitrary cosmic-ray composition models. The simulation covered zenith angles 0 – 70$^{\circ}$ and full azimuthal range (0 – 360$^{\circ}$). The detector's response is modeled using G4WCDA within the Geant4 framework ~\cite{SA03}. The simulated events are subsequently weighted according to some composition model. To determine the systematic uncertainties due to the choice of composition model, we have applied the Gaisser~\cite{GAS13}, H\"orandel~\cite{HOR04}, Global Spline Fit (GSF)~\cite{GSF18}, and Gaisser-Stanev-Tilav (GST) ~\cite{GST14} models. For all subsequent analyses, we consistently categorize the primary cosmic rays into five distinct groups: protons, helium, CNO, MgAlSi, and iron. 

	\subsubsection{Event Quality Cuts}
	To ensure that the selected events are well reconstructed, the following event selection criteria are applied on the simulated as well as data events:
			
	\begin{enumerate}
		\item The number of cells fired $(N_{\text{cell}})$ is greater than 120.
		\item The shower core should lie at least 30 m inside the WCDA boundary.
		\item Zenith angle $\theta<45 ^\circ $.
    \end{enumerate}
    
    The event quality cut $(N_{\text{cell}}>120)$ is selected to ensure that the angular resolution of the Moon shadow as estimated from the Gaussian fit of the projected deficit map along the declination for the first $N_{\text{pe}}$ bin is 0.393$^\circ $ or less (corresponding to $\sigma_{\text{PSF}}=0.372^\circ $ after correcting for the radius of the Moon's disk), as indicated by the value of $\sigma_\text{DEC}$ in the last column of Table~\ref{1_qcr100}. This is slightly better than the angular resolution 0.48$^\circ$, the benchmark established in~\cite{CRAB26} for $100<N_{\text{cell}}<200$. A further increase in $N_{\text{cell}}$ cut slightly improves angular resolution at the expense of statistical error. Therefore, with the chosen range, we aim to strike a balance between statistical error and reconstruction quality. The choice of containing the shower core at least 30 m inside the edge is based on our MC study, in which we calculate the shifts along the RA for different values of core selection. We find that the shifts nearly saturate at the chosen value.

    \renewcommand{\arraystretch}{1.5}
	\begin{table*}[htb]
		\begin{center}
      
			\caption{Left to Right: $N_{pe}$ ranges, Moon-shadow minimum significance, Moon shadow shifts along RA ($\alpha$) and DEC ($\delta$) with statistical errors, and Gaussian widths $\sigma_{\text{RA}}$. and $\sigma_{\text{DC}}$.}
			\label{1_qcr100}
				\begin{tabular}{cccccccccc}
					\hline
					\hline
					
					&  \textbf{No. } &  \textbf{$N_{pe}$} &  \textbf{Minimum}   &   \textbf{$\Delta^{\circ}  (\text{RA}) $} &  \textbf{$\Delta^{\circ} (\text{DEC}) $} &  \textbf {$\sigma_{RA} (^{\circ})$} &  \textbf {$\sigma_{DEC} (^{\circ})$} & \\
					
					& \textbf{} & \textbf{Ranges} &  \textbf{Significance}  & \textbf{} & \textbf & &\\
					\hline
					\hline
					& 1 &  1000 - 1390  &  -43 &   $-0.524 \pm (0.013)_{\text{stat}}$ &  $-0.037 \pm (0.010)_{\text{stat}}$ &  $0.515 \pm 0.014$ &  $0.393 \pm 0.011$ \\	
					\hline
					&  2 &  1390 - 1930 &  -44 &  $-0.414 \pm (0.012)_{\text{stat}}$ &  $-0.035 \pm (0.009)_{\text{stat}}$ &  $0.466 \pm 0.013 $ &  $0.352 \pm 0.010$ \\
					\hline
					&  3 &  1930 - 2683 &  -44 &  $-0.319 \pm (0.010)_{\text{stat}}$ &  $-0.025 \pm (0.008)_{\text{stat}}$ &  $0.414 \pm 0.011 $ &   $0.314 \pm 0.009$\\
					\hline
					&  4 &  2683 - 3728 &  -43 &  $-0.245 \pm (0.009)_{\text{stat}}$ &  $-0.026 \pm (0.008)_{\text{stat}}$ &  $0.357 \pm 0.010 $ &  $0.293 \pm 0.008$\\
					\hline
					&  5 &  3728 - 5180 &  -42 &  $-0.188 \pm (0.009)_{\text{stat}} $&  $-0.013 \pm (0.007)_{\text{stat}}$ &  $0.317 \pm 0.009 $ & $0.259 \pm 0.008$ \\
					\hline
					&  6 &  5180 - 7197  &  -37 &   $-0.139 \pm (0.010)_{\text{stat}}$ &   $0.007 \pm (0.008)_{\text{stat}}$ &  $0.316 \pm 0.010$ &  $0.252 \pm 0.009$\\
					\hline
					& 7 &  7197 - 10000 &  -31 &  $-0.087 \pm (0.011)_{\text{stat}}$ &  $0.006 \pm (0.009)_{\text{stat}}$ &  $0.316 \pm 0.012 $ &  $0.247 \pm 0.010$\\
					\hline
					&  8 &  10000 - 13895 &  -26 &  $-0.063 \pm (0.012)_{\text{stat}}$ &  $0.039 \pm (0.011)_{\text{stat}}$ &  $0.272 \pm 0.013$  & $0.262 \pm 0.013$ \\
					\hline					
				\end{tabular}
		\end{center}
	\end{table*}

\renewcommand{\arraystretch}{1.5}
	\begin{table*}[htb]
		\begin{center}      
			\caption{Values of $\sigma(^\circ)$ of the PSF for each bin, obtained from WCDA-MC.}
			\label{2_psf_ra}
				\begin{tabular}{cccccccc}
                    \hline
					\hline
                    &  \textbf{No.} & \textbf{Proton}         & \textbf{Helium} & \textbf{CNO}  & \textbf{MgAlSi} & \textbf{Fe }  & \\
                    \hline
					\hline
					& 1 &  0.365 &  0.455 & 0.562 & 0.604 & 0.727 &\\	
					\hline
					& 2 &  0.330 &  0.415 & 0.508 & 0.591 & 0.662 &\\
					\hline
					& 3 &  0.302 &  0.373 & 0.454 & 0.545 & 0.568 &\\
					\hline
					& 4 &  0.267 &  0.336 & 0.409 & 0.496 & 0.536 &\\
					\hline
					& 5 &  0.254 &  0.298 & 0.376 & 0.461 & 0.468 &\\
					\hline
					& 6 &  0.221 &  0.284 & 0.385 & 0.374 & 0.627 &\\
					\hline
					& 7 &  0.219 &  0.246 & 0.352 & 0.361 & 0.417 &\\
					\hline
					& 8 &  0.198 &  0.239 & 0.334 & 0.348 & 0.352 &\\
					\hline
					
				\end{tabular}
		\end{center}
	\end{table*}
    
	%

			
	\subsubsection{Energy Estimator and MC shifts}		
\begin{figure}[th]
				\begin{center}
					\includegraphics[width=0.8\linewidth]{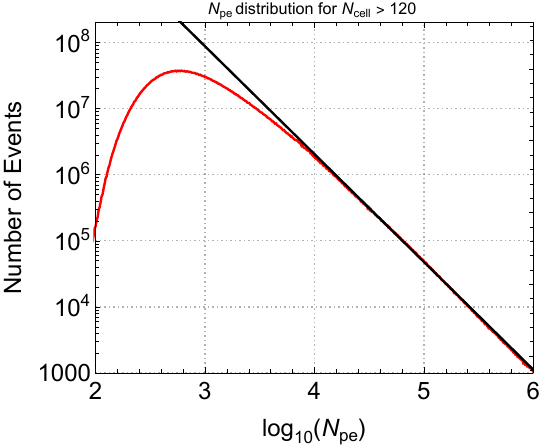}
				\end{center}
                \caption[ms]{Distribution of the total number of photoelectrons $N_\text{pe}$ (red) recorded in the radius of 20 to 120 meters from the shower core. The black solid line corresponds to a power-law fit of form $a N_{pe}^{b}$, with fitted values $a=12.84$ and $b=-1.63$.}            
				\label{2_npe}
			\end{figure}

			The number of photoelectrons $(N_{\text{pe}})$ detected by photomultiplier tubes (PMTs) residing in water Cherenkov detectors serves as a crucial observable for estimating the primary energy of cosmic rays. It corresponds to the total amount of light detected and is thus a good estimator of the energy deposited in the water. In this work, the number of photoelectrons within a radius of 20 to 120 meters from the shower core and within 30 ns from the shower conical front are specifically used as the energy estimator to avoid the effect of saturation near the shower core. The distribution of $N_{\text{pe}}$ along with power-law fit are shown in FIG.~\ref{2_npe}. 

\begin{figure}[th]
				\begin{center}
					\includegraphics[width=0.95\linewidth]{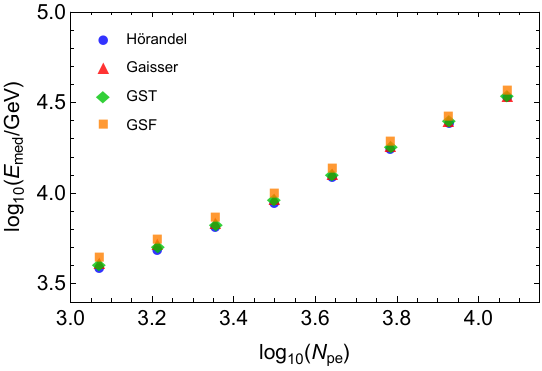}
				\end{center}
				\caption[ER]{Median energy ($E_\text{med}$) vs $N_\text{pe}$ for four composition models (H\"orandel, Gaisser, GSF and GST) including the contribution of five groups of primary cosmic-ray particles.}
				\label{3_mc_en}
			\end{figure}

            To study the response of the estimator using WCDA MC, we divided $N_{\text{pe}}$ into eight equally spaced bins on a logarithmic scale, as listed in Table~\ref{1_qcr100}. For each bin, we estimate the median value from the corresponding true energy histogram, including five distinct primary cosmic-ray groups (proton, helium, CNO, MgAlSi, and Fe). The FIG.~\ref{3_mc_en} presents the relation between the median energy $E_{med}$ and the estimator $N_{\text{pe}}$ for four composition models (H\"orandel, Gaisser, GSF and GST). These values are related through a power law, which on the logarithmic scale is given as follows:   
            \begin{equation}\label{en}
				\log_{10} {\left(\dfrac{E}{\text{GeV}}\right)} = p_0 + p_1 \log_{10} (N_{\text{pe}}).
			\end{equation}
            The fitted values of the parameters $p_{0}$ and $p_{1}$ for four different composition models (H\"orandel, Gaisser, GSF and GST) are listed in Table~\ref{2_model}.

            \begin{table}[hth]
	           \begin{center}
		          \caption{The fitted values of the parameters $p_{0}$ and $p_{1}$ for four composition models.}
		          \begin{tabular}{ccccc}
			      \hline
			      \hline	
                   &   \textbf{Composition } & \textbf {$p_{0}$}  &  {$p_{1}$}  \\
                     &  \textbf{Model} &   &    \\
			     \hline
			     \hline
			     &  Gaisser     & 0.504 &  0.992 \\
                 &              H\"orandel  & 0.407 &  1.014 \\
                 &              GST         & 0.452 &  1.004 \\
                 &              GSF         & 0.550 &  0.985 \\
			     \hline
		         \end{tabular}  
                  \label{2_model}
	             \end{center}
            \end{table}



    
       \subsection{Ray-Tracing (RT) Simulations in the Geomagnetic Field}     
\begin{figure*}[htb]
		  \begin{center}
			\includegraphics[width=0.8\linewidth]{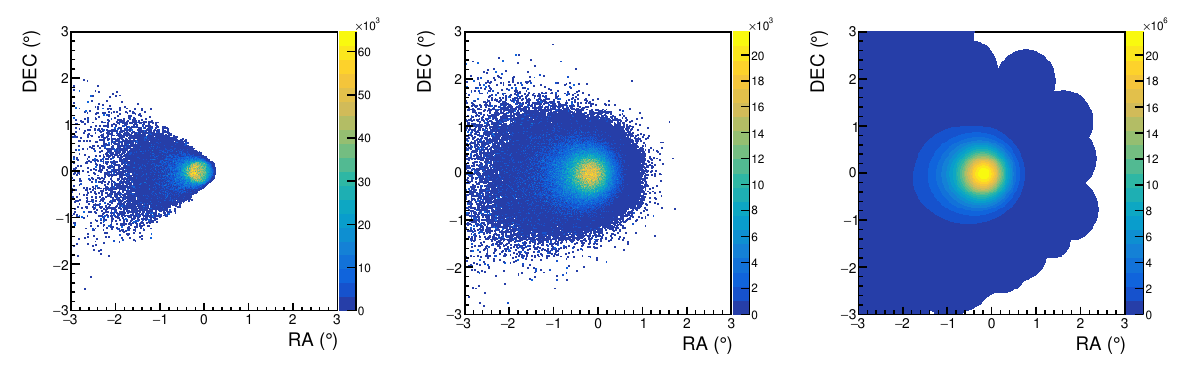}
		      \end{center}
		      \caption[1D]{2D deficit maps obtained by ray-tracing for all $N_\text{pe}$ bins. Left: GMF only, assuming an ideal detector with no PSF. Middle: Combined effect of the detector PSF and the GMF. Right: Result after applying a smoothing procedure to PSF-corrected map. The color scale indicates the number of showers per pixel.}
		  \label{4_RT}
	    \end{figure*}

	       To estimate the westward shift of the Moon’s shadow, an accurate ray-tracing simulation is required, taking into account the magnetic fields that govern the Earth–Moon system. In this region, the geomagnetic field (GMF) dominates, while the interplanetary magnetic field (IMF) carried by solar wind is subdominant and neglected in this work, and the Moon's magnetic field is negligible. A simplified dipole model of the GMF provides a first-order estimate of the total deflection in the Moon's shadow ~\cite{SI11}. This estimate, valid for nearly vertical cosmic-ray primaries with energies above a few TeV, is given by:
       \begin{equation}\label{del}
		  \Delta^{\circ} =  -1.59^{\circ} \dfrac{Z} {E(\text{TeV})}
	   \end{equation}
       
       where $Z$ is the atomic number and $E$ is the energy of the primary particle in $\text{TeV}$. To ensure an accurate GMF simulation, we adopted the Tsyganenko-IGRF (T-IGRF) model with a K$_{\text{p}}$ index of 2, a parameter that characterizes global geomagnetic disturbance and drives the external field component \cite{NAT95}. This model accounts for both internal geomagnetic field and external magnetospheric effects using empirical data collected from direct observations. To estimate the shift for each $N_\text{pe}$ bin, we used WCDA MC events that meet the selection criteria specified in Section III–A-1. We applied a ray-tracing method \cite{BBB11}, using the true energy and direction of the primary cosmic ray particles. To capture the complete range of the Moon's apparent trajectory and seasonal variations, we executed the ray-tracing simulation over a full one-year period, with a time resolution of 10 seconds. We generated events within a circular window of the $5^{\circ}$ radius centered on the Moon, and only those events that were actually blocked by the Moon’s disk were used for further analysis. A clear westward shift caused by the GMF is evident in the resulting two-dimensional (2D) map of deficit counts for all $N_{\text{pe}}$ bins, as shown in the left panel of FIG.~\ref{4_RT}.

\begin{figure}[th]
			\begin{center}
                \includegraphics[width=0.95\linewidth]{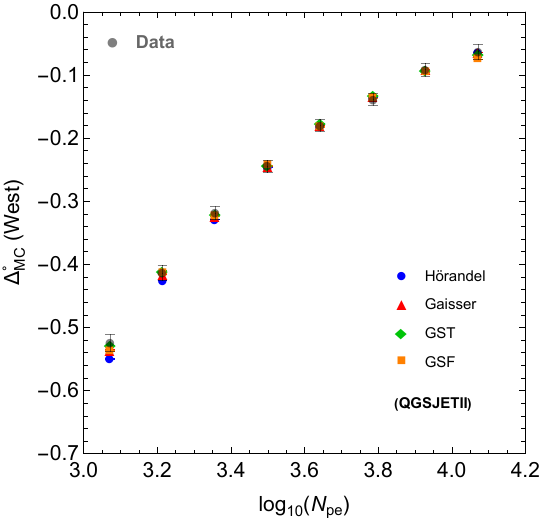}
			\end{center}
			\caption[mS]{Moon shadow shifts along RA from the experimental data (black color) compared with the MC values for four composition models:  H\"orandel, Gaisser, GSF, and GST. MC values include the contributions of five groups of primary cosmic ray particles.}
			\label{5_mc_shift}
		\end{figure}

        To account for the detector’s finite angular resolution, we applied a Gaussian PSF to the raw deficit map. The $\sigma$ values of the PSF along the RA and DEC directions, listed in Table~\ref{2_psf_ra}, were obtained from WCDA MC simulations for each bin, incorporating the contributions from five primary cosmic-ray particle groups. The resulting 2D map after applying the PSF is shown in the middle panel of FIG.~\ref{4_RT}. To ensure direct comparability with the Moon shadow analysis, the map is also subjected to the same smoothing procedure used for the significance map in the Moon shadow analysis (right panel of FIG.~\ref{4_RT}). To extract numerical values for the shifts along RA, we projected the 2D PSF-convolved maps along RA and fitted each with a 1D Gaussian function. These shifts, including the contribution of the five primary cosmic-ray groups, are shown in FIG.~\ref{5_mc_shift} for four composition models: H\"orandel, Gaisser, GSF, and GST.
      
        \par We use the values of the estimated shift in degrees and define $Z_{\text{eff}}$ for each $N_{\text{pe}}$ bin, in terms of the product of the shift ($\Delta^{\circ}_{\text{RT}}$) and the median energy ($E_{\text{med}}$) as follows.  
        \begin{equation}\label{del_mean}		
            Z_{\text{eff}} = \frac{-\Delta^{\circ}_{\text{RT}}E_{\text{med}}(\text{TeV})}{1.59^{\circ}}
		\end{equation} 

\begin{figure}[th]
			\begin{center}
			\includegraphics[width=0.95\linewidth]{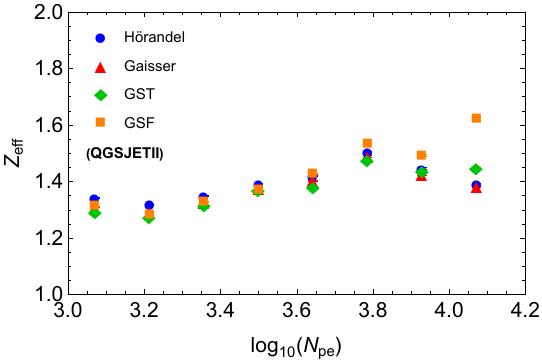}
			\end{center}
			\caption[mz01]{The averaged values of $ {Z_{\text{eff}}}$ for four different composition models (H\"orandel, Gaisser, GSF, GST). The values include the contributions of five groups of primary cosmic ray particles.}
			\label{6_mc_MZ}
		\end{figure}

        In FIG.~\ref{6_mc_MZ}, we show the values of $Z_{\text{eff}}$ for four composition models. In the second column of Table~\ref{3_energy}, we list the corresponding mean values along with statistical and systematic errors. The systematic error is estimated through the standard deviation of the values from four composition models for each $N_\text{pe}$ bin. The mean values of $Z_{\text{eff}}$ are used to determine the calibrated energy as follows:

        \begin{equation}
            \label{del2}
			E_{\text{cal}}(\text{TeV}) =  -\dfrac{{Z}_{\text{eff}}} {\Delta^{\circ}_{\text{data}}} \times 1.59^{\circ}
		\end{equation}	
		where $\Delta^{\circ}_{\text{data}}$ is the westward shift for each bin measured from WCDA data. 
        

\subsection{Analysis of the Moon shadow displacement in East-West direction}
	\subsubsection{Data Selection}
	In this work, we analyze the experimental data spanning from March 17, 2021, to July 31, 2024 (3.4 years) recorded by the WCDA full-array to observe the Moon shadow. All events collected during this period are processed using the same selection criteria defined earlier. The total effective observation time of the Moon with zenith angles smaller than $45^\circ$ is 4269.64 hours, and the total number of events is about 1.33 billion with arrival direction within 5° of the Moon's position. 
    
    \par
\begin{figure}[th]
		\begin{center}
			\includegraphics[width=0.95\linewidth]{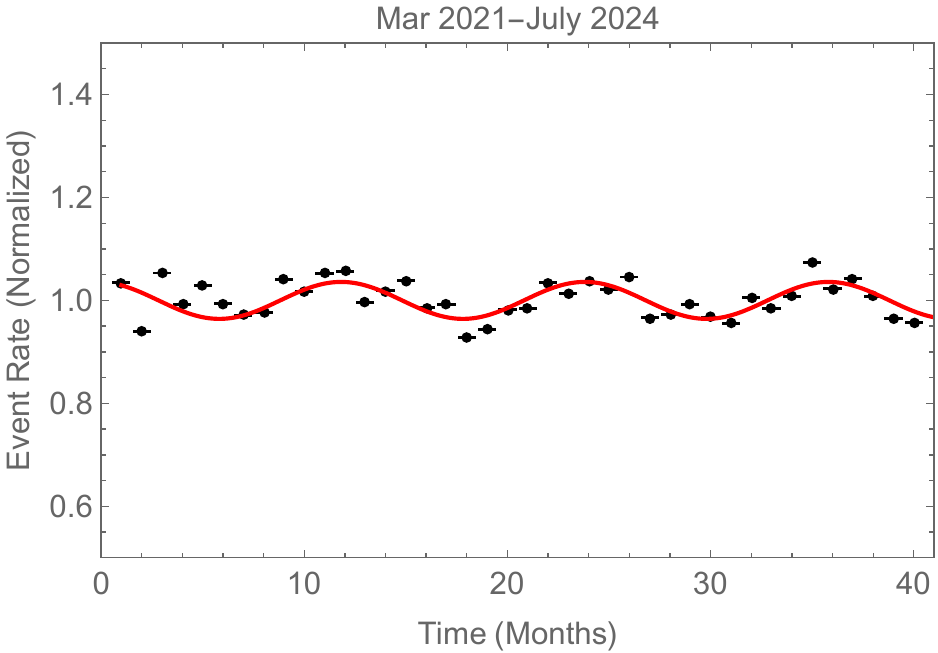}
		\end{center}
		\caption [rate]{Detector stability profile from March 2021 to July 2024. Each point represents the Moon's visibility period for one month (normalized to the mean of total events with data selection cuts).}
		\label{7_Erate}
	\end{figure}

 FIG.~\ref{7_Erate} presents the profile of the monthly event rate for the on-source window centered on the Moon's position over the period of 3.4 years. The overall standard deviation of the normalized event rate is 0.036. In particular, there is a seasonal modulation, as evident by a sinusoidal fit with a one-year periodicity. The standard deviation relative to the fitted curve (red) is 0.025, which can be attributed to other factors, such as variation in water clarity and variation in the Moon's zenith angle.
        

    \begin{figure*}[htb]
		\centering
		\begin{center}
			\includegraphics[width=0.45\linewidth]{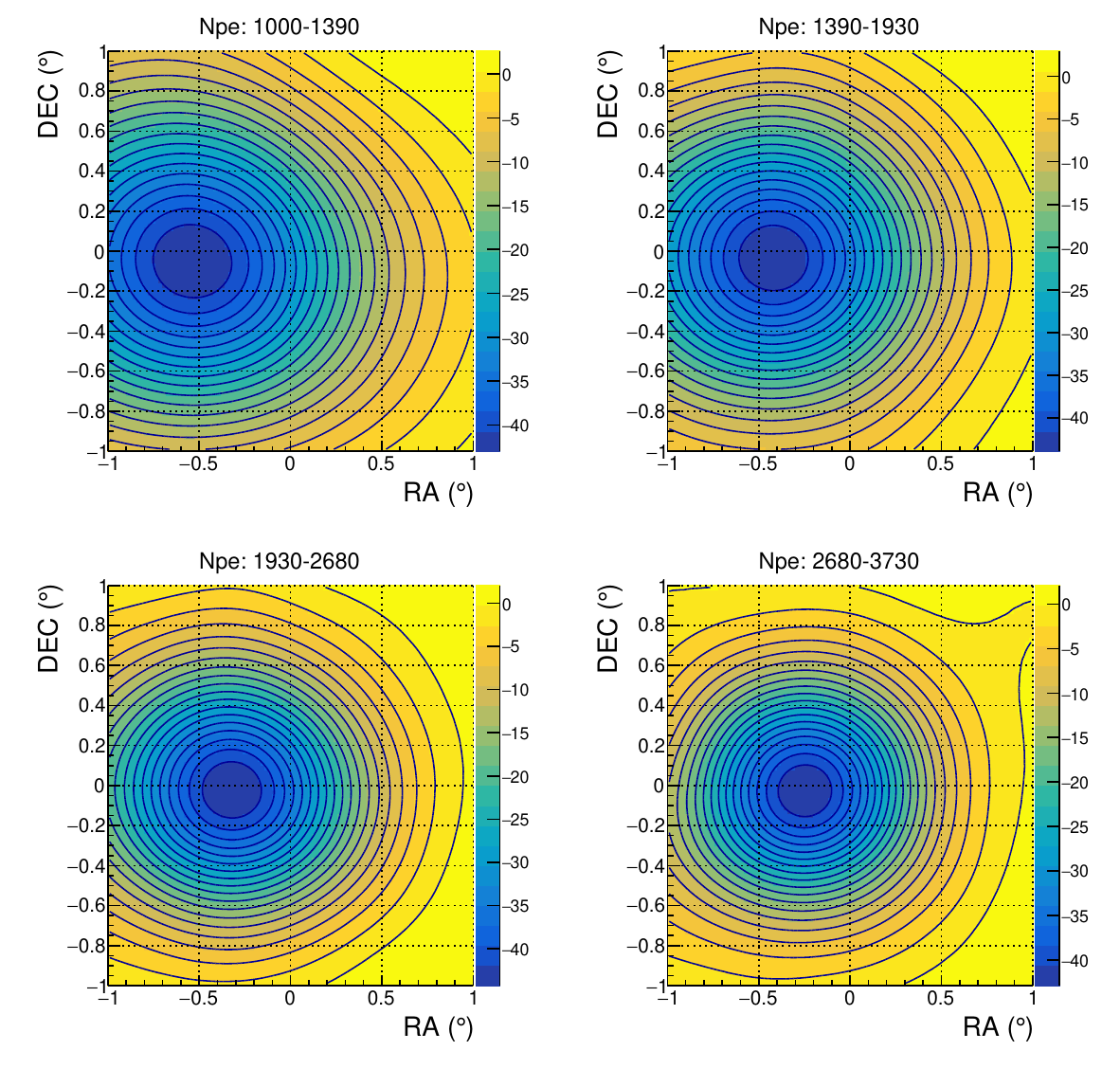}
			\includegraphics[width=0.45\linewidth]{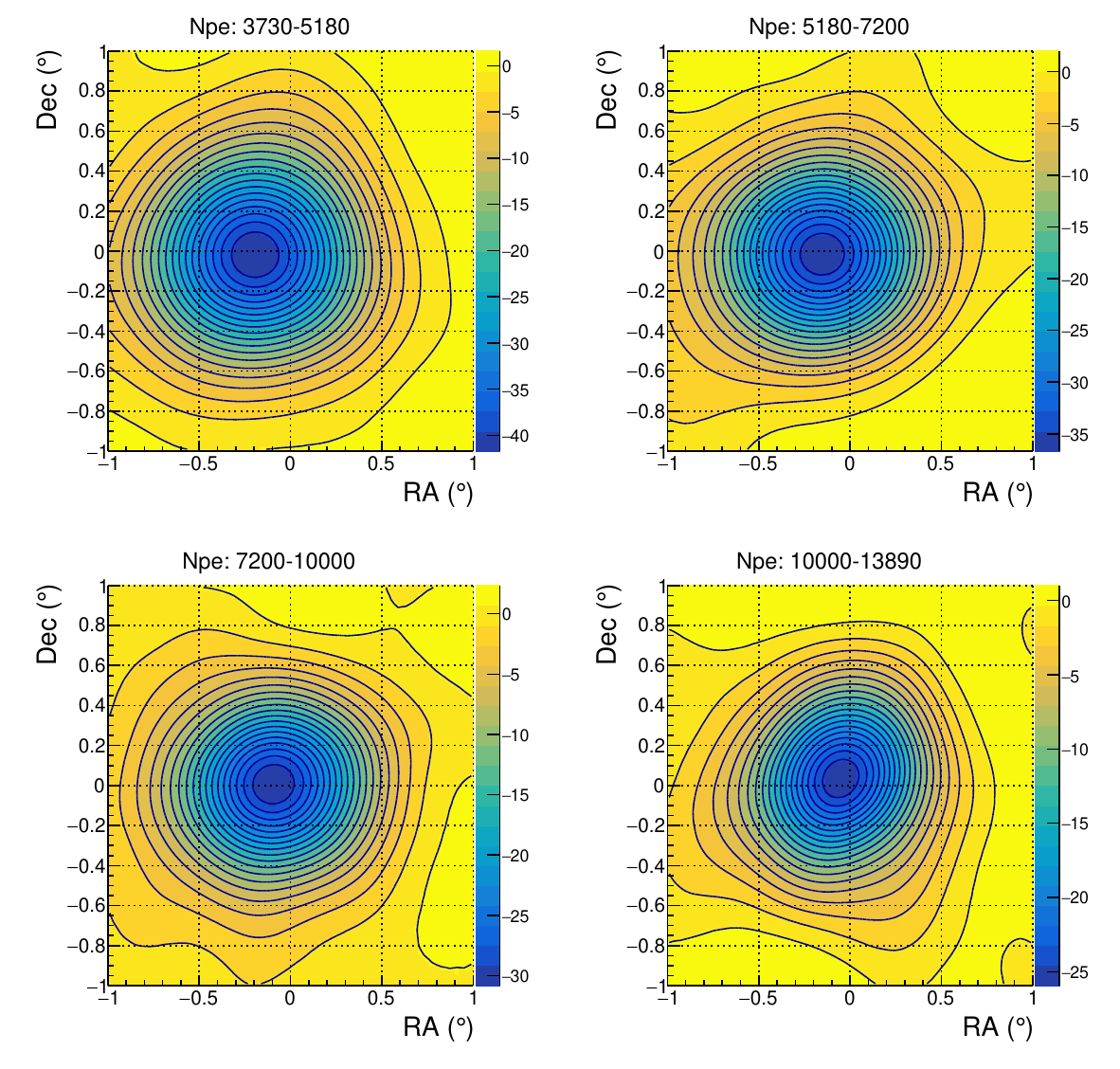}
		\end{center}
		\caption[NPE Dist.]{From left to right: 2D smoothed significance maps of the Moon shadow for each bin, constructed from events satisfying the selection cuts given in section III-A-I.}
		\label{8_2DMaps}
	\end{figure*}
    



    \begin{figure*}[htb]
		\begin{center}
			\includegraphics[width=0.8\linewidth]{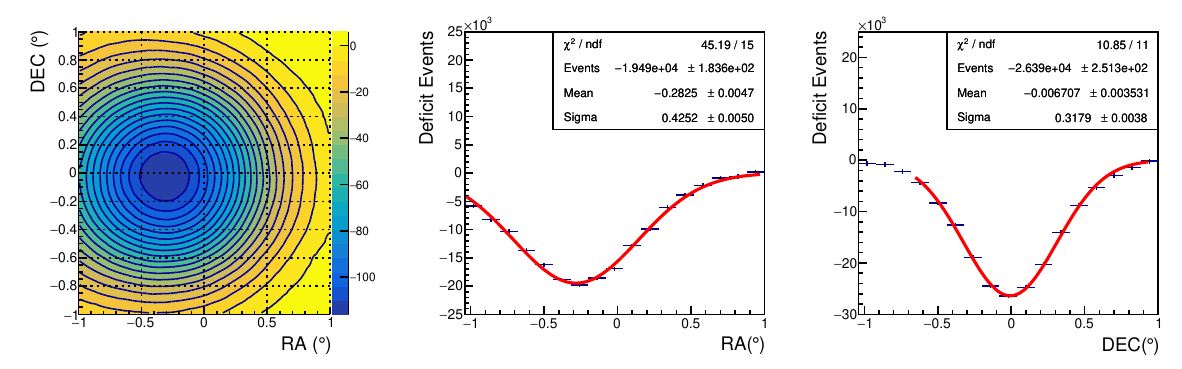}
		\end{center}
		\caption[1D]{(Left) 2D-smoothed significance map for selected events satisfying the event quality cuts described in section III-A-I, observed by LHAASO-WCDA for the period March 2021 - July 2024.  (Middle) The shift along RA caused by the Moon shadow. (Right) The shift along the DEC direction.}
		\label{9_1D}
	\end{figure*}

\subsubsection{Measurement of Moon Shadow Displacement in RA and DEC Directions}   
    
	\par To measure the energy-dependent shift in the Moon’s shadow, the events are grouped into eight $N_{\text{pe}}$ bins, as specified in the second column of Table~\ref{1_qcr100}. For each bin, we produce three circular maps in celestial coordinates ($\alpha$, $\delta$): an event map centered on the Moon’s position, a background map to estimate the cosmic-ray deficit, and a significance map to evaluate statistical significance. In the Moon shadow study, an accurate background estimation is crucial for detecting the Moon’s shadow deficit against the isotropic cosmic-ray background. The background map is constructed using the equi-zenith method, with ten off-source windows symmetrically placed around the on-source window at the same zenith angle. The radius of both the on-source and the off-source windows is taken $5^{\circ}$. With this size, the windows cover the entire azimuthal circle at the zenith angle of $18^{\circ}$. To avoid event duplication below it, the size of the windows is dynamically adjusted to ensure the exclusive counting of each event. To construct the significance map, we divide the event and background maps into bins of size $0.02^{\circ} \times 0.02^{\circ}$ and compute the statistical significance of each bin using the Li and Ma formula~\cite{LM83}. To enhance the signal-to-background ratio and produce a clear picture of the cosmic-ray Moon shadow, we smoothed the significance maps using a Gaussian weight. The resulting maps for the eight $N_{\text{pe}}$ bins, whose ranges are specified in the second column of Table~\ref{1_qcr100}, are shown in FIG.~\ref{8_2DMaps}. In FIG.~\ref{9_1D} (left), we show the combined map for all events that meet the event selection criteria described in Section~\ref{sec:selection}.
   
    \begin{figure}[th]
		\begin{center}
			\includegraphics[width=0.95\linewidth]{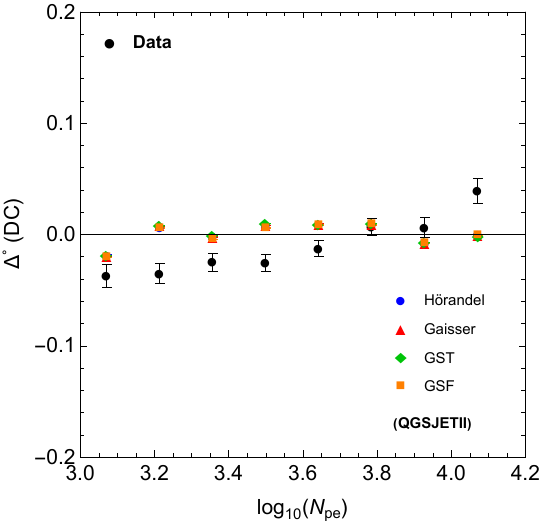}
		\end{center}
		\caption[shdc]{The measured shifts along DC from the experimental data (black color) compared with the MC values for four composition models:  H\"orandel, Gaisser, GSF, and GST.}
		\label{10_dc}
	\end{figure}

    \par We determine the shifts in the east–west (RA) and north–south (DEC) directions by taking 1D projections of the deficit maps along the corresponding axes. Each projection is fitted with a 1D-Gaussian function $A_0 \exp[-(x-x_0)^2/(2\sigma^2)]$, where $x_{0}$ and $\sigma$ correspond to the displacement and width of the Moon shadow. In the middle and right panels of FIG.~\ref{9_1D}, we explicitly show the projected graphs for RA and DEC, along with 1D Gaussian fits, for all events satisfying the event selection criteria. The fitted values of the shifts along RA and DEC for each $N_\text{pe}$ bin are reported in Table~\ref{1_qcr100}. The shifts along RA are plotted in FIG.~\ref{5_mc_shift} together with the corresponding MC values for four composition models for comparison, and those along DEC are shown in FIG.~\ref{10_dc}. The vertical error bars attached to the data points indicate the uncertainties resulting from the statistical error. The plot in FIG.~\ref{5_mc_shift} shows that the Moon's shadow is predominantly displaced along RA, exhibiting an energy dependence as predicted by ray-tracing, while the displacement along DEC changing monotonically from -0.04 to +0.04, indicates the limited pointing accuracy of the detector. We used these measured non-zero shifts along DEC to estimate the systematic error in shifts along RA due to pointing accuracy and propagate it to calibrated values of energies. 
\subsubsection{Determination of the Energy Scale of LHAASO-WCDA}
	This study aims to calibrate the energy scale of WCDA using the cosmic ray Moon shadow. The Moon shadow technique provides a standard approach for calibrating the energy scale of cosmic ray detectors by exploiting the deflection of charged particles caused by the GMF. The quantitative measurement of the Moon shadow's displacement, where higher-energy particles experience less deflection, enables us to calibrate the energy scale of primary cosmic rays. The procedure consists of the following steps: 

	\begin{enumerate}
		\item Measure the westward displacement of the Moon's shadow for each bin of an energy estimator, which in our case is $N_\text{pe}$.

        \item Using detailed shower and detector simulations (i.e., CORSIKA and detector response models), determine the median energy $E_\text{med}$ for each $N_{\text{pe}}$ bin.
	
		\item Calculate the expected Moon shadow shifts by ray-tracing ($\Delta_{\text{RT}}$) of cosmic rays for a specific composition and GMF for each $N_{\text{pe}}$ bin.
		
		\item Compute effective $Z_{\text{eff}}$ by combining $\Delta_\text{MC}$ with $E_\text{med}$ for each $N_\text{pe}$ bin using Eq. \ref{del_mean}.  
        
		
		\item Finally, determine the calibrated energy $E_{\text{cal}}$ by combining the measured Moon shadow shifts $\Delta_{\text{data}}$ and the effective value $Z_{\text{eff}}$ for each $N_\text{pe}$ bin using Eq. \ref{del2}.
		\end{enumerate}
	
\section{Determination of the primary energy using the Moon shadow displacement}   

\begin{figure}[th]
			\begin{center}
            \includegraphics[width=0.96\linewidth]{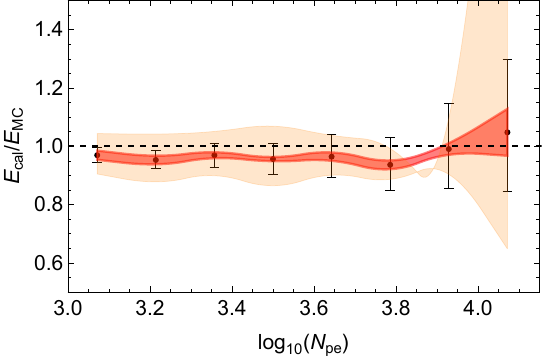}
			\end{center}
			\caption[EC1]{Ratio $E_\text{cal}/E_\text{MC}$ for each $N_{\text{pe}}$ bin and corresponding statistical errors are shown by black points and attached black bars, respectively. Light-orange and dark-orange bands represent systematic uncertainties due to pointing accuracy and composition model, respectively.}
			\label{11_EC}
		\end{figure}

    The calibrated values of energy $E_{\text{cal}}$ from the Moon shadow data and corresponding errors for each $N_{\text{pe}}$ bin are listed in Table~\ref{3_energy}. The first and second asymmetric errors in $E_\text{cal}$ are the statistical error and the systematic error due the pointing accuracy, respectively, whereas the symmetric error is systematic due to the composition model. Moreover, we present the ratio $E_\text{cal}/E_\text{MC}$ in FIG.~\ref{11_EC}, where $E_\text{MC}$ is the composition-model averaged energy derived from the MC simulation for each bin. 
    The comparison of calibrated and MC values on linear scale provides a clear view of deviations. The error bars in the plot correspond to the statistical error, while the light- and dark-orange bands represent the systematic errors due to the pointing accuracy and composition model, respectively. 
  
    \renewcommand{\arraystretch}{1.5}
	\begin{table*}[htb]
		\begin{center}
			\caption{Left to Right: bin number, composition-model averaged MC value of $Z_{\text{eff}}$ with statistical error and systematic error due to composition model, Moon shadow shift along RA with statistical error, composition-model averaged MC median energy ${E}_\text{med}$ with statistical error and systematic error due to composition model, and calibrated energies ${E}_{\text{cal}}$ in TeV. The first and second asymmetric errors in $E_\text{cal}$ are statistical and systematic due to pointing accuracy, respectively, whereas the symmetric error is the systematic due to the composition model.}
		
			
			\label{3_energy}
				\begin{tabular}{ccccccc}
				    \hline
				    \hline
			
				    & \textbf{} & \textbf{Composition-model}  &   \textbf{} &   \textbf {Composition-model} &  \textbf{Calibrated } &\\
				    &  \textbf{Bins No.} &  \textbf{averaged value}  &   \textbf{$\Delta^{\circ}  (\text{RA)}$}  &    \textbf {averaged MC median} &  \textbf{Energy } &\\
				
				& \textbf{} &  \textbf{$Z_{\text{eff}}$}  &   \textbf{} &   \textbf {energy $E_{\text{med}}(\text{TeV})$} &   \textbf{$E_{\text{cal}} (\text{TeV})$}&\\
				\hline
				\hline
				& 1 & $1.312 \pm (0.002)_{\text{stat}} \pm (0.021)_{\text{sys}}$ &   $-0.524 \pm (0.013)_{\text{stat}}$ & $4.11 \pm (0.0020)_{\text{stat}} \pm (0.021)_{\text{sys}} $&  $ 3.98^{+0.09 + 0.26}_{-0.10 - 0.30} \pm (0.065)_{\left<Z\right>} $ \\
				\hline
				& 2 & $1.291 \pm (0.002)_{\text{stat}}  \pm (0.021)_{\text{sys}}$ & $-0.414 \pm (0.012)_{\text{stat}}$  &  $5.12 \pm (0.0022)_{\text{stat}} \pm (0.020)_{\text{sys}} $&  $ 4.95^{+0.16 + 0.39}_{-0.16 - 0.46} \pm (0.077)_{\left<Z\right>} $  \\ 
				\hline
				& 3 & $1.330 \pm (0.002)_{\text{stat}}  \pm (0.014)_{\text{sys}}$ & $-0.319 \pm (0.010)_{\text{stat}}$  & $ 6.83 \pm (0.0023)_{\text{stat}} \pm (0.019)_{\text{sys}} $ & $6.62^{+0.27 + 0.48}_{-0.29 - 0.56} \pm (0.070)_{\left<Z\right>} $ \\
				\hline
				&4 & $1.376 \pm (0.003)_{\text{stat}}  \pm (0.010)_{\text{sys}} $ & $-0.245 \pm (0.009)_{\text{stat}} $ & $ 9.34 \pm (0.0028)_{\text{stat}} \pm (0.018)_{\text{sys}} $ & $8.93^{+0.46 + 0.85}_{-0.52 - 1.05} \pm (0.068)_{\left<Z\right>}  $ \\
				\hline
				& 5 & $1.407 \pm (0.004)_{\text{stat}}  \pm (0.020)_{\text{sys}}$ & $-0.188 \pm (0.009)_{\text{stat}}$ & $ 12.90 \pm (0.0031)_{\text{stat}} \pm (0.015)_{\text{sys}} $& $11.93^{+0.80 + 0.75}_{-0.92 - 0.86} \pm (0.177)_{\left<Z\right>} $ \\
				\hline
				& 6 & $1.499 \pm (0.005)_{\text{stat}}  \pm (0.023)_{\text{sys}}$ &  $-0.139 \pm (0.010)_{\text{stat}}$ &  $ 18.30 \pm (0.0035)_{\text{stat}} \pm (0.014)_{\text{sys}} $& $17.13^{+1.50 + 0.91}_{-1.83 - 1.01} \pm (0.275)_{\left<Z\right>}$  \\
				\hline
				& 7 & $1.450 \pm (0.007)_{\text{stat}}  \pm (0.029)_{\text{sys}}$ & $-0.087 \pm (0.011)_{\text{stat}}$ & $25.29 \pm (0.0040)_{\text{stat}}\pm (0.011)_{\text{sys}} $ & $26.42^{+3.52 + 2.43}_{-4.81 - 2.98} \pm (0.533)_{\left<Z\right>}$ \\
				\hline
				& 8 & $1.459 \pm (0.008)_{\text{stat}}  \pm (0.111)_{\text{sys}}$ & $-0.063 \pm (0.012)_{\text{stat}}$ & $34.95 \pm (0.0050)_{\text{stat}} \pm (0.011)_{\text{sys}} $ & $36.60^{+6.40 + 13.96}_{-9.85 - 58.95} \pm (2.811)_{\left<Z\right>} $  \\
				\hline
			
			\end{tabular}
		\end{center}
	\end{table*}
 
	\par
\begin{figure}[th]
			\begin{center}
				\includegraphics[width=0.95\linewidth]{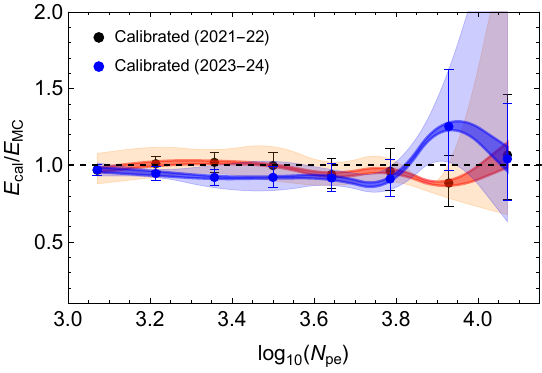}
			\end{center}
			\caption[rate]{Ratio $E_\text{cal}/E_\text{MC}$ for different time periods.}
			\label{12_YEC}
		\end{figure}

\begin{figure}[th]
			\begin{center}
				\includegraphics[width=0.95\linewidth]{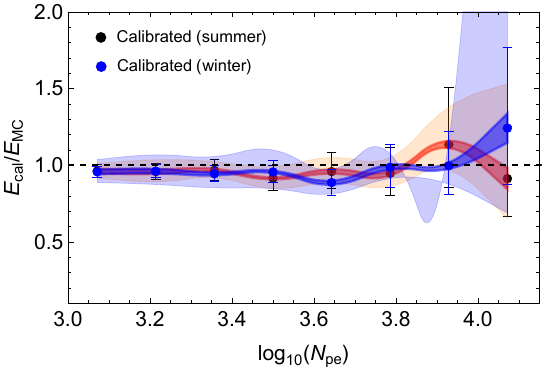}
			\end{center}
			\caption[rate]{Ratio $E_\text{cal}/E_\text{MC}$ for different seasons.}
			\label{13_SEC}
		\end{figure}

 Additionally, the ratios $E_\text{cal}/E_\text{MC}$ obtained for different years and seasons are presented in FIG.\ref{12_YEC} and FIG.\ref{13_SEC} respectively. In FIG.\ref{12_YEC}, $E_\text{cal}/E_\text{MC}$ values for the data collected during 2021-2022 and 2023-24 are denoted by black and blue points, respectively, with the corresponding error bars representing the statistical uncertainties. The light-orange and light-blue bands indicate the systematic uncertainty due to pointing accuracy for 2021–2022 and 2023–2024, respectively, while the orange and blue regions represent the systematic uncertainty due to the composition model for the respective periods. A seasonal comparison of the $E_\text{cal}/E_\text{MC}$ ratios is shown in FIG. \ref{13_SEC}. The black and blue points correspond to summer (May-September) and winter (October-April), respectively. The color scheme used to draw the shades of systematic uncertainties is the same as that used in FIG.\ref{12_YEC}. The plots show that the WCDA detector exhibits excellent temporal and seasonal stability. To quantitatively evaluate the agreement between the calibrated and MC values, we introduce a scale shift parameter $\epsilon$ defined through the relation

    \begin{equation}\label{sf}
			E_{\text{cal}} = (1+\epsilon) E_\text{{med}}
	\end{equation}
    This parameter measures the fractional deviation between the calibrated energy scale and the model-dependent simulated scale. We perform a ($\chi^{2}$) minimization to determine the best-fit value of $\epsilon = -0.04 \pm  0.03 $, taking into account both statistical and systematic errors in $E_\text{cal}$ and $E_\text{med}$. The corresponding 95\% confidence interval for $\epsilon$, in the range 4-37 TeV, is $-9\%$ to $+0.2\%$. This tightly constrained interval bounds the maximum allowable systematic deviation of the simulated energy scale. 

	\section{Conclusion}	
	Accurately determining primary cosmic ray energy is fundamental to experimental astroparticle physics. This study presents a comprehensive energy calibration of the LHAASO-WCDA detector using the Moon shadow technique, demonstrating its excellent pointing accuracy and angular resolution for cosmic ray observations.  

     The number of photoelectrons $N_\text{pe}$ recorded by PMTs serves as an effective energy estimator and a reliable proxy for the total energy deposited by secondary particles. This estimator provides an independent, model-supported approach to validate and refine the energy scale of large air shower arrays. We measured the westward displacement of the Moon's shadow caused by the geomagnetic field (GMF) and found that it decreases with increasing energy, consistent with MC simulation. Detailed detector simulation is employed to relate the observed Moon shadow shifts to the primary energy and the effective atomic number factor of the incoming particles. This approach allows us to establish a direct link between the $N_\text{pe}$ measured on the ground and the median energy of the cosmic ray primaries.
	 
	In summary, combining Moon shadow displacement with $N_\text{pe}$ based energy estimation creates a comprehensive calibration framework that strengthens confidence in the measured energy spectra. This work is vital for studying spectral features like the knee, ankle, and possible cutoffs that provide clues about acceleration mechanisms, propagation effects, and origins. By validating the detector’s energy scale, the frontiers of high-energy cosmic ray observations and the long-standing mysteries of cosmic ray origins can be further explored.

    \section{Acknowledgments}
    We would like to thank the staff members of the National Facility for Large High Altitude Air Shower Observatory (LHAASO) (https://cstr.cn/31117.02.LHAASO) who work at the LHAASO site above 4400 meter above sea level year round to maintain the detector and keep the water recycling system, electricity power supply and other components of the experiment operating smoothly. We are grateful to Chengdu Management Committee of Tianfu New Area for the constant financial support for the research with LHAASO data. We appreciate the computing and data service support provided by the National High Energy Physics Data Center for the data analysis in this paper. This work was supported by the National Key R\&D Program of China (Grant No. 2025YFE0202600); The National Natural Science Foundation of China No.12393851, No.12393852, No.12393853, No.12393854, (if using KM2A data: NSFC No.12205314, No.12105301, No.12305120, No.12261160362, No.12105294, No.U1931201, No.12375107), (if using WCDA data: NSFC No.12173039, the Department of Science and Technology of Sichuan Province, China No.24NSFSC2319, Project for Young Scientists in Basic Research of Chinese Academy of Sciences No.YSBR-061) (if using WFCTA data: NSFC No.12275280, No.11905240, No.12375106, No.12261141691, Innovation Project of IHEP No. E25451U2, Sichuan Province Science Foundation for Distinguished Young Scholars No.2022JDJQ0043, the Youth Innovation Promotion Association of the Chinese Academy of Sciences (CAS YIPA) (Grant No. 2023019)) (if using meteorological data: We are grateful to Institute of Plateau Meteorology, CMA Chengdu to maintain meteorological data), and in Thailand from the NSRF via the Research and Innovation Acceleration Agency for Competitiveness and Area Development (RCAD) (Program Management Unit for Technology and Innovation for Future Industries (PMU-B) : Brainpower for Future Industries) [grant number B39G690003].

\bibliographystyle{apsrev4-2}  
\bibliography{references}      
\end{document}